\documentclass[aps,pra,twocolumn,groupedaddress,showpacs]{revtex4-1}

\usepackage{amssymb,amsfonts,amsmath}
\usepackage{bm}
\usepackage{graphicx}

\begin{document}

% ===== \newcommand ==================================================

% bra- and ket-vectors
\newcommand{\bra}[1]{\left< #1 \right|}
\newcommand{\ket}[1]{\left| #1 \right>}
\newcommand{\bracket}[2]{\left< \left. #1 \right| #2 \right>}
\newcommand{\Bracket}[3]{\left< #1 \left| #2 \right| #3 \right>}
\newcommand{\Erwartungswert}[1]{\left< #1 \right>}
\newcommand{\FT}[1]{ \mathcal{F} \left\lbrace  #1 \right\rbrace  }
\newcommand{\IFT}[1]{\mathcal{F}^{-1} \left\lbrace  #1 \right\rbrace  }

% norm und absoute value
\newcommand{\norm}[1]{\| #1 \|}
\newcommand{\abs}[1]{\left| #1 \right|}

% definition of single letters
\newcommand{\p}{\partial}
\newcommand{\R}{\vec{r}}
\newcommand{\laplace}{\Delta}
\newcommand{\Det}{\text{Det}}
\newcommand{\Emf}{E_\text{mf}}
\newcommand{\ii}{\mathrm{i}}
\newcommand{\dd}{\mathrm{d}}
\newcommand{\Ns}{N_{\! s}}

\newcommand{\Arj}{A_{\! \rho\! j}}		\newcommand{\Arl}{A_{\! \rho l}}
\newcommand{\Azj}{A_{\! z\! j}}			\newcommand{\Azl}{A_{\! z l}}

\newcommand{\Arrj}{A_{\! \rho\! j}^{\! r}}	\newcommand{\Arrl}{A_{\! \rho\! l}^{\! r}}
\newcommand{\Arzj}{A_{\! z\! j}^{\! r}}		\newcommand{\Arzl}{A_{\! z\! l}^{\! r}}

\newcommand{\Airj}{A_{\! \rho\! j}^{\! i}}	\newcommand{\Airl}{A_{\! \rho\! l}^{\! i}}
\newcommand{\Aizj}{A_{\! z\! j}^{\! i}}		\newcommand{\Aizl}{A_{\! z\! l}^{\! i}}

\newcommand{\Aixj}{A_{\! x\! j}^{\! i}}
\newcommand{\Aiyj}{A_{\! y\! j}^{\! i}}
\newcommand{\Axj}{A_{\! x\! j}}
\newcommand{\Ayj}{A_{\! y\! j}}

\newcommand{\alpharjl}{\alpha_{\! jl}^{\rho}}
\newcommand{\alphazjl}{\alpha_{\! jl}^{z}}

\newcommand{\qrj}{q_{\rho j}}
\newcommand{\qzj}{q_{z j}}
\newcommand{\prj}{p_{\! \rho j}}
\newcommand{\pzj}{p_{\! z j}}
\newcommand{\qrl}{q_{\rho l}}
\newcommand{\qzl}{q_{z l}}

% vectors
\renewcommand{\vec}[1]{\bm{#1}}

\title{Variational calculations on multilayer stacks of dipolar Bose-Einstein condensates}

\author{Andrej Junginger}
\author{J\"org Main}
\author{G\"unter Wunner}

\affiliation{Institut f\"ur Theoretische Physik 1, Universit\"at Stuttgart, 70550 Stuttgart, Germany}

\date{\today}

\begin{abstract}
We investigate a multilayer stack of dipolar Bose-Einstein condensates in terms of a simple Gaussian variational  ansatz and demonstrate that this arrangement is characterized by the existence several stationary states. Using a Hamiltonian picture we show that in an excited stack there is a coupled motion of the individual condensates by which they exchange energy. We find that for high excitations the interaction between the single condensates can induce the collapse of one of them. We furthermore demonstrate that one collapse in the stack can force other collapses, too. We discuss the possibility of experimentally observing the coupled motion and the relevance of the variational results found to full numerical investigations.
\end{abstract}


\maketitle

\section{Introduction}
Over the past fifteen years Bose-Einstein condensates (BECs) have become an active field of theoretical and experimental investigations. The experimental realization of dipolar BECs with chromium atoms \cite{Pfau2005} (for a recent review and further references see Ref. \cite{Review2009}) has opened the way to actually observe the theoretically predicted phenomena of radial and angular rotons, anisotropic solitons, biconcave shapes of the ground state, etc. \cite{Ronen2007, Tik2008, Dutta2007} which should exist in such condensates with an additional long-range interaction. Moreover, dipolar BECs are of special importance since the interactions between the atoms can be tuned from predominantly short-range to the dominance of the long-range dipole-dipole interaction (DDI) by manipulating the $s$--wave scattering length via Feshbach resonances.

\begin{figure}[t]
\includegraphics[width=0.47\columnwidth]{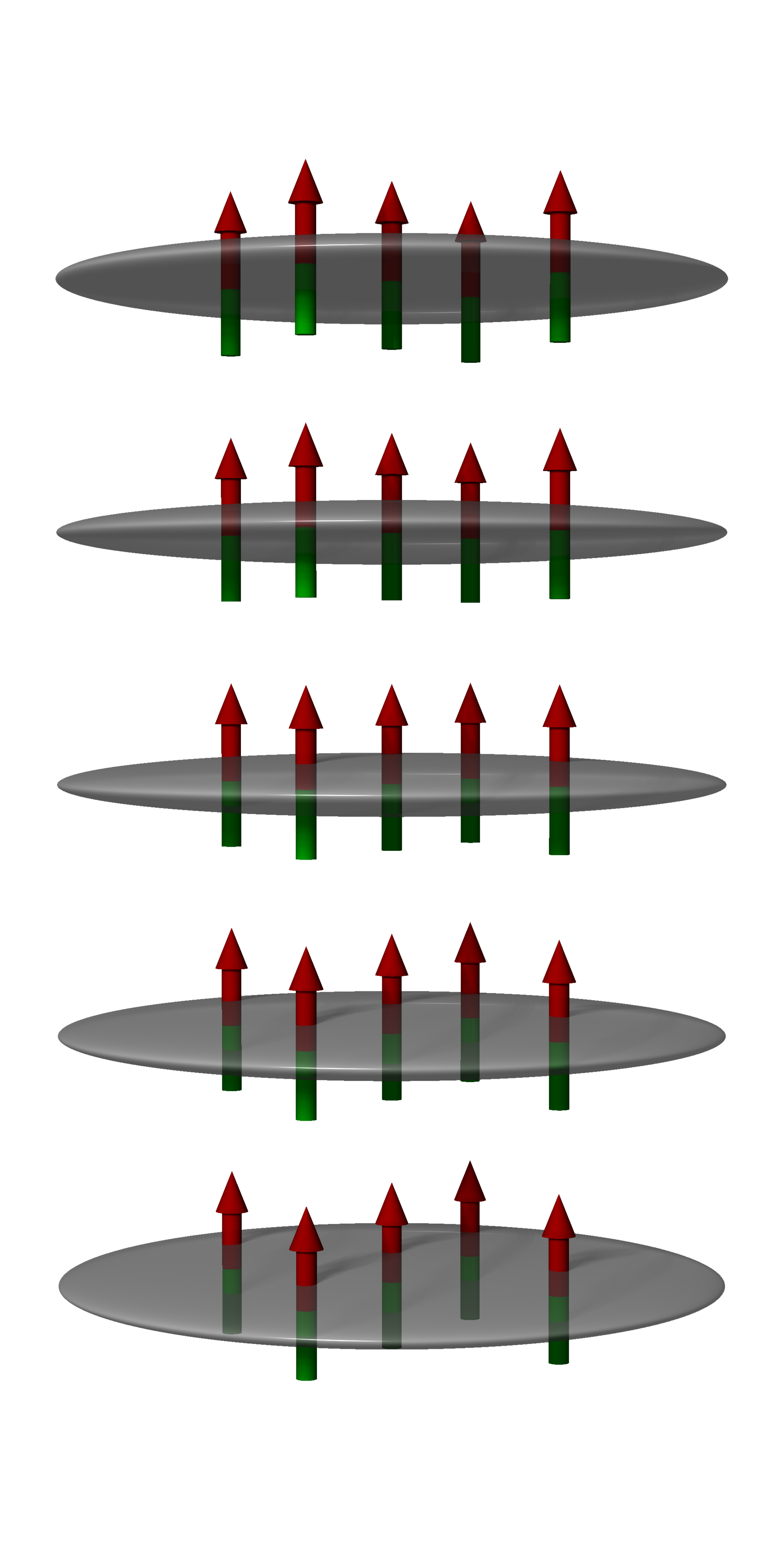}
 \caption{A multilayer stack of dipolar BECs, each placed in a trap that is assumed to be very oblate. The single condensates are displaced by a distance $\Delta$, respectively, and coupled by the long-range dipole-dipole interaction (cf. \cite{KlawunnSantosPRA}).}
 \label{Fig-Stack3D}
\end{figure}

Recently, multilayer stacks of dipolar BECs have been in the focus of theoretical investigations \cite{KlawunnSantosPRA, koeberle2009}, even though they have not yet been realized experimentally. In Refs. \cite{KlawunnSantosPRA, koeberle2009} numerically exact calculations on grids have been performed via imaginary time evolution, revealing structured ground state wave functions and the roton instability. By contrast, the purpose of this paper is to investigate such a multilayer stack of dipolar BECs in the framework of a variational approach with a Gaussian type orbital for each layer. In this way we shall not be able to catch exotic features of dipolar BECs such as structured wave functions or the roton instability but we can determine different stationary states and study in particular the dynamics of the coupled BECs. Furthermore, we show that excited BECs exchange energy and that the scattering length as well as the distance between the BECs have a strong influence on this energy exchange. For highly excited BECs this energy exchange can induce the collapse of one of the BECs, and the collapse of a single BEC in the stack can cause other collapses, too.

\section{Theory}
In accordance with Refs. \cite{koeberle2009, KlawunnSantosPRA}, we investigate a stack of $\Ns$ dipolar BECs. Each of the condensates is arranged in an axisymmetric trap that is assumed to be very oblate (see Fig. \ref{Fig-Stack3D}) and the traps (and consequently the BECs) are displaced by a distance $\Delta$ from each other in the $z$--direction.

At very low temperatures the quantum gas in each condensate $j$ can be described by a single wave function $\psi_j(\R,t)$ whose dynamics obeys the Gross-Pitaevskii equation (GPE)
\begin{align}
 	\biggl( - \Delta  + \hat{V}_\text{trap} + \hat{V}_\text{c}  + \hat{V}_\text{d} \biggr) \psi_j(\R,t) = \ii \p_t \psi_j (\R,t), \nonumber
\end{align} where
\begin{equation*}
 	\hat{V}_\text{trap} = N^4\gamma_\rho^2 \rho^2 + N^4\gamma_z^2 \left( z+ \frac{(\Ns+1-2j)\Delta}{2} \right)^2
\end{equation*} is the potential energy caused by the harmonic traps each condensate is placed in. These traps are arranged in the coordinate system in a way that the $z=0$--plane is always at the center of the stack. Here the GPE is written in dimensionless form obtained by introducing ``atomic'' units \cite{koeberle2009NJP}: Using the mass $m$ of the bosons and their magnetic moment $\mu$, we define a  ``dipole length'' $a_d = m\mu_0\mu^2/(2\pi\hbar^2)$, a unit energy $E_d = \hbar^2/(2ma_d^2)$ and a unit frequency $\omega_d = E_d/\hbar$. The trap geometry is determined by a mean trap frequency $\bar{\omega} = (\omega_\rho^2 \omega_z)^{1/3}$ and a trap aspect ratio $\lambda=\omega_z / \omega_\rho$. The parameters $\gamma_{\rho, z}$ in the GPE are connected to the trap frequencies by $\gamma_{\rho, z} = \omega_{\rho,z} / (2\omega_d)$. Furthermore, we assume each BEC to consist of the same number of particles $N$ so that we can apply a particle number scaling $\vec{r} \to N a_d \vec{r}$ and $E \to E_d E /N^2$ to make the interaction potentials independent of the particle number. The term
\begin{equation*}
 	\hat{V}_\text{c} = 8\pi \frac{a}{a_d} \abs{\psi_j (\R)}^2
\end{equation*} represents the contact interaction, which depends on the $s$--wave scattering length $a$, measured in units of $a_d$, and 
\begin{equation*}
 	\hat{V}_\text{d} = \int \! \dd^3 r' ~ \frac{1 - 3 \frac{({z} - {z}')^2}{(\R - \R')^2}}{ \abs{\R - \R'}^3 } ~ \sum_{l=1}^{\Ns} \abs{  \psi_l (\R')}^2
\end{equation*} is the potential energy caused by the dipole-dipole interaction (DDI) between all the atoms. The long-range nature of the DDI demands the summation $l=1{,} \ldots {,}\Ns$ and couples all the individual condensates.

We investigate the multilayer stack of dipolar BECs variationally using a cylindrically symmetric Gaussian trial wave function 
\begin{equation*}
 	\psi(\rho,z,t)  = \sum_{j=1}^{\Ns} \psi_{\!j} (\rho,z,t),
\end{equation*} where each condensate $j$ is described by a single wave function
\begin{multline*}
 	\psi_{\!j} (\rho, z,t) = \left( \frac{2 \Airj}{ {\pi}} \right)^{1/2} \left( \frac{2 \Aizj }{ {\pi}} \right)^{1/4} \times \\  \exp \left( \mathrm{i} A_{\rho\! j} \rho^2  + \mathrm{i} A_{z\! j} \left(z+ \frac{(\Ns+1-2j)\Delta}{2}  \right)^2 \right).  \label{Gl-NormierteWellenfunktion}
\end{multline*} Here $\Arj = \Arrj + \ii \Airj$ and $\Azj = \Arzj + \ii \Airj$ are complex valued and time-dependent width parameters that are split into their real and imaginary parts and have to satisfy $\Airj, \Aizj >0$. The wave functions $\psi_j (\vec{r})$ are assumed to be non-overlapping and normalized:
\begin{equation*}
 	\int\! \dd^3r~ \abs{\psi_j(\R)}^2 = 1.
\end{equation*} We have to choose the value of $\Delta$ in such a way that on the one hand it is large enough so that we can ensure the wave functions to be non-overlapping and on the other hand that it is small enough to make the interaction between the BECs sufficiently large.

To investigate the stack of multilayer BECs, we will first calculate the mean-field energy and then apply a time-dependent variational principle. The dynamics of the system will be calculated using an equivalent Hamiltonian picture which reveals the physics of the system in a more transparent way than that of complex width parameters.

\subsection{Calculation of the mean-field energy}

The mean-field energy $\Emf$ of the arrangement is given by 
\begin{align*} 
	E_\text{mf} &= \int\! \dd^3r~ \psi^* (\R) \biggl( - \Delta + {V}_\text{trap}  + \frac{1}{2} {V}_\text{c} + \frac{1}{2} {V}_\text{dip} \biggr) \psi (\R) 
\end{align*} where, as usual, we have to insert a factor of $1/2$ for the contact and the dipole-dipole interaction to avoid a double counting of the two-particle interactions. The contributions of the kinetic energy, the potential energy in the harmonic traps and the contact interaction can easily be calculated:
\begin{align}
	\int\! \dd^3r~ {\psi^* (\R)} (-\Delta ){\psi (\R)}  &=  \sum_{j=1}^{\Ns}~ 2\Airj + 2\frac{[\Arrj]^2}{\Airj} \nonumber \\ & \qquad \qquad + \Aizj + \frac{[\Arzj]^2}{\Aizj}\; ,\label{eq-Ekin} \\
	\int\! \dd^3r~ {\psi^* (\R)}{{V}_\text{trap}}{\psi(\R)} &= \sum_{j=1}^{\Ns} ~ \frac{N^4 \gamma_\rho^2}{2 \Airj} + \frac{N^4 \gamma_z^2}{4 \Aizj}\; , \label{eq-Etrap} \\
	 \frac{1}{2} \int\! \dd^3r~ {\psi^* (\R)}{  {V}_\text{c}}{\psi(\R)} &= \sum_{j=1}^{\Ns}~ \frac{4}{\sqrt{\pi}} \frac{a}{a_d} \Airj \sqrt{\Aizj} .\label{eq-Escat}
\end{align} All integrals occurring here are elementary. The contribution of the DDI, given by 
\begin{equation*}
 \frac{1}{2} \int\! \dd^3r~ {\psi^*(\R)}{ V_\text{d}}{\psi(\R)} ,
\end{equation*} is more difficult to calculate. We do this Fourier transforming the DDI potential 
\begin{align*}
	\hat{V}_\text{d} &= \IFT{ \FT{\int\! \dd^3r' ~\frac{1 - 3 \frac{({z} - {z}')^2}{(\R - \R')^2}}{ \abs{\R - \R'}^3 } \sum_{l=1}^{\Ns} \abs{\psi_l(\R')}^2 } } \\
	&= \IFT{   (2\pi)^{3/2} \cdot \FT{\frac{1 - 3 \frac{({z} - {z}')^2}{(\R - \R')^2}}{ \abs{\R - \R'}^3 } }    \times \right. \\
	& \qquad \qquad  \qquad \qquad \qquad \qquad \left. \FT{\sum_{l=1}^{\Ns} \abs{\psi_l(\R')}^2 } }.
\end{align*} where $\FT{...}$ and $\IFT{...}$ denote the symmetrical form of the Fourier transform and its inverse. In the last step, we have applied the convolution theorem. For the first Fourier transform one obtains \cite{GoralSantosPRA}
\medskip

\begin{align*}
 	(2\pi)^{3/2} \cdot \FT{\frac{1 - 3 \frac{({z} - {z}')^2}{(\R - \R')^2}}{ \abs{\R - \R'}^3 } } = \frac{4 \pi}{3}\left( \frac{3k_z^2}{\vec{k}^2 } -1 \right).
\end{align*} 
\bigskip

The Fourier transform of $\sum_l \abs{\psi_l(\R)}^2$ can be calculated in a straightforward way, and the result is
\begin{align*}
	&\FT{ \sum_{l=1}^{\Ns} \abs{\psi_l(\R)}^2 } = \frac{1}{(2\pi)^{3/2}} \times \\
	& \qquad \sum_{l=1}^{\Ns}  \exp \left( -\frac{k_\rho^2} {8A_{\rho l}^i} - \frac{k_z^2}{8A_{z l}^i} -  \mathrm{i} \frac{ (\Ns+1-2l)\Delta k_z }{2} \right) ,
\end{align*} where the term $- \mathrm{i} (\Ns+1-2l)\Delta k_z /2$ occurs because of the displacement of the individual condensates in the $z$--direction. We can now write the contribution of the DDI to the variational mean-field energy as
\begin{widetext}
\begin{multline}
	\frac{1}{2} \int\! \dd^3r~ {\psi^*(\R)}{ \hat{V}_\text{d}}{\psi(\R)} =  \sum_{l,j=1}^{\Ns}  \int \dd^3r~  \abs{\psi_j(\R)}^2 \times \\  \mathcal{F}^{-1} \Biggl\lbrace \frac{4\pi}{3 (2\pi)^{3/2}} \left( \frac{3k_z^2}	{\vec{k}^2 } -1 \right) \cdot \exp  \left( -\frac{k_\rho^2} {8A_{\rho l}^i} - \frac{k_z^2}{8A_{z l}^i } -  \mathrm{i} \frac{ (\Ns+1-2l)\Delta k_z }{2} \right) \Biggr\rbrace  
\end{multline}
% \end{align*}
% \end{widetext} 
where we first integrate over $\vec{r}$ and afterwards solve the integral over $\vec{k}$ resulting from the inverse Fourier transform. The variational dipole interaction energy finally reads
% \begin{widetext}
\begin{align}
 	\frac{1}{2} \int\! \dd^3r~ {\psi^*(\R)}{ {V}_\text{d}}{\psi(\R)} =&~ \frac{1}{8 \sqrt{\pi}}  \sum_{\substack{j,l=1 \\ j \neq l}}^{\Ns} \int_1^\infty \frac{\dd s}{s} \exp\left(- \frac{(j-l)^2 	\Delta^2}{4g}\right) \left( \frac{1}{g^{3/2}} - \frac{(j-l)^2\Delta^2}{2g^{5/2}} \right) \nonumber \\
     	 & -\frac{4}{3} \sqrt{\frac{2}{\pi}} \sum_{j{,}l=1}^{\Ns} \frac{1}{ \alpharjl\sqrt{\alphazjl}} \exp \left( -\frac{2 (j-l)^2\Delta^2}{\alphazjl}\right)  , \label{eq-Edip2}
\end{align} 
\end{widetext}
where we have introduced the function $g=\alphazjl/8 + 2 \alpharjl (s-1) $ and new variables $\alpha_{\! jl}^{\rho,z} = [ A_{\rho,z j}^i ]^{-1} + [ A_{\rho,z l}^i ]^{-1}$ for brevity. It is useful to split the sum into a part $j=l$ and a part $j \neq l$ (i.e. a term that describes the DDI between all the atoms belonging to the same condensate and a term that describes the interaction between different condensates) because the remaining integral can be solved analytically for $j=l$. In this case the result is
\begin{multline}
 \frac{1}{2}\int\! \dd^3r~ {\psi^*(\R)}{ V_\text{dip}^{{(l=j)}}}{\psi(\R) } =  \\ 
 \frac{2}{3\sqrt{\pi}} \sum_{j=1}^{\Ns} \frac{\Airj \sqrt{\Aizj}}{\eta_j -1}  \left( 1 + 2 \eta_j - \frac{3 \eta_j \arctan \sqrt{\eta_j -1}}{\sqrt{\eta_j -1}} \right) \label{eq-Edip1}
\end{multline} with $ \eta_j = \Aizj / \Airj $. The mean-field energy of the whole stack is then given by subsuming  Eqs. \eqref{eq-Ekin} - \eqref{eq-Edip1} with the constraint $j \neq l$ for the summation in Eq. \eqref{eq-Edip2}.

\subsection{Dynamics of multilayer stacks of dipolar BECs}

To describe the dynamics of the stacked BECs, we apply a time-dependent variational principle (TDVP) to solve the time-dependent GPE
\begin{gather*}
 	\ii \dot{\psi} (t) = H \psi(t) 
\end{gather*} with the parameter-dependent wave function $\psi(t) = \psi( \vec{A}(t) )$ where $\vec{A} = ( \vec{A}_{\rho}, \vec{A}_z )^t$ is the set of time-dependent variational parameters.
The TDVP has already been applied to condensates with an attractive 
gravity-like $1/r$-interaction \cite{Cartarius2008} and to dipolar BECs
\cite{koeberle2009NJP}.
The application of the McLachlan variational principle \cite{TDVPMcLachlan} 
leads to a set of first order differential equations, describing the 
time evolution of the real and imaginary parts of the width parameters.

Since it is more descriptive, we present the dynamics of the arrangement using an equivalent Hamiltonian picture, describing the wave function of each condensate by a particle moving in a $2\Ns$-dimensional space. The Hamiltonian form is obtained introducing generalized coordinates
\begin{align*}
 \qrj = \sqrt{ \langle \rho^2 \rangle } = \frac{1}{\sqrt{2 \Airj}},\\  \qzj = \sqrt{ \langle z^2 \rangle } =\frac{1}{2\sqrt{\Aizj}}
\end{align*} and generalized momenta defined by 
\begin{align*}
 \prj = \frac{\sqrt{2} \Arrj}{\sqrt{\Airj}}, \quad \pzj = \frac{\Arzj}{\sqrt{\Aizj}}.
\end{align*} With this definition, the generalized coordinates can directly be interpreted as the ``extension'' of the condensate in the radial and $z$--direction, respectively. The Hamiltonian is formally obtained by substituting 
\begin{equation*}
 \left( \Airj, \Arrj,\Aizj, \Arzj  \right)  \longrightarrow  \left( \qrj, \qzj, \prj, \pzj \right)
\end{equation*} in the mean-field energy, which results in the Hamiltonian $H=T+V$, where
\begin{equation*}
 	T = \sum_{j=1}^{\Ns}  \prj^2 + \pzj^2
\end{equation*}  is the kinetic energy and
\begin{widetext}
\begin{align}
 	V =&~ \sum_{j=1}^{\Ns} \frac{1}{\qrj^2} + \frac{1}{4\qzj^2} + N^4\gamma_\rho^2 \qrj^2 + N^4\gamma_z^2 \qzj^2
	 + \frac{1}{\sqrt{\pi}} \frac{a}{a_d} \frac{1}{\qrj^2 \qzj} + \frac{1}{6\sqrt{\pi}} \frac{1 + 2 \eta_j - 3\eta_j \arctan \sqrt{\eta_j - 1}}{\qrj^2 \qzj  \sqrt{\eta_j-1}} \nonumber \\
	& + \frac{1}{8\sqrt{\pi}}\sum_{\substack{j,l=1 \\ j \neq l}}^{\Ns} \int_1^\infty \frac{\dd s}{s} \left( \frac{1}{g^{3/2}} - \frac{(j-l)^2\Delta^2}{2g^{5/2}} \right) \exp \left(-\frac{(j-l)^2\Delta^2}{4g}\right)   \nonumber \\
	& - \frac{1}{3} \sqrt{\frac{2}{\pi}} \sum_{\substack{j,l=1 \\ j \neq l}}^{\Ns} \frac{ 1 }{(\qrj^2 + \qrl^2) \sqrt{\qzj^2 + \qzl^2}} \exp \left( -\frac{(j-l)^2 \Delta^2}{2(\qrj^2 + \qrl^2)}  \right) \label{eq-Vext}
\end{align}
\end{widetext} is the external potential in which the ``particle'' moves. Here the substitution leads to $g=(\qzl^2+\qzl^2)/2 + 4(\qrj^2+\qrl^2)(s-1)$. %Increasing $\Delta$ the value of the integral as well as the exponential function in the second summand decrease, so we can scale the strength of the particle interaction by varying $\Delta$. 

The time evolution of the particle in the $2\Ns$-dimensional space $(\vec{q}_\rho, \vec{q}_z)$ is then determined by the Hamiltonian equations of motion
\begin{equation}
 	\dot{q}_{\rho,z j} = \frac{\p H}{\p p_{\rho,z j}} \qquad  \text{and} \qquad \dot{p}_{\rho,z j} = - \frac{\p H}{\p q_{\rho,zj}}, \label{eq-Ham-eq-motion}
\end{equation} which, after backward substituting  $\left(\qrj, \qzj, \prj, \pzj \right) \longrightarrow \left(\Airj, \Arrj,\Aizj, \Arzj \right)$, yield the same equations of motion obtained by applying the TDVP, hence the Hamiltonian form is equivalent to describing the condensates using parameter-dependent Gaussian trial wave functions.

Summarizing all the coordinates $\vec{q} = (\vec{q}_\rho, \vec{q}_z)^t$ in one vector, the Hamiltonian equations of motion can easily be brought into the form
\begin{equation}
	 \ddot{\vec{q}} = \vec{f}(\vec{q}) \label{eq-q2punkt}
\end{equation} where $ \vec{f}(\vec{q}) = -2\p V / \p \vec{q}$ is a function of all the coordinates and the external parameters. Depending on the value of $a/a_d$ and the number $\Ns$ of BECs in the stack, $V$ is characterized by one minimum and several saddle points, each corresponding to a stationary state of the multilayer stack of BECs, given by the fixed points
\begin{align*}
 \dot{\vec{q}} = 0, \quad  \ddot{\vec{q}} = 0.
\end{align*}

To investigate the stability of the different states  and the motion in the very vicinity of their fixed points, we follow the usual procedure and linearize Eq. \eqref{eq-q2punkt} around one of its fixed points $\vec{q}_0$, which results in the differential equation
\begin{equation}
 \ddot{\vec{u}} = J\vec{u}. \label{eq-linearization}
\end{equation} Here the vector $\vec{u} = \vec{q} - \vec{q}_0$ denotes the deviation of the particle from the fixed point and  $\left. J=\p \vec{f}(\vec{q})/\p \vec{q} \right|_{\vec{q}=\vec{q}_0}$  is the Jacobian matrix. Eq. \eqref{eq-linearization} is that of a system of $2\Ns$ coupled oscillators and can simply be solved using the ansatz $\vec{u} = \vec{u}_0 e^{\kappa t}$ with a complex parameter $\kappa$. Inserting this ansatz into the differential equation yields the eigenvalue equation
\begin{equation}
 \left(J - \kappa^2 \right)  \vec{u}_0 = 0
\end{equation} where $\kappa^2$ are the eigenvalues of the Jacobian matrix. It can easily be shown, that $J$ is symmetric, consequently the eigenvalues are purely real. 

If all the eigenvalues are negative, $\vec{u}$ oscillates around the fixed point according to $ \vec{u} \sim e^{i \omega t} $ with a frequency $ \omega = \sqrt{- \kappa^2} $ and the fixed point is stable. If one of the eigenvalues is positive, there is a contribution $\vec{u} \sim e^{\kappa t}$ and the fixed point is unstable.

\section{Results}

\subsection{Stationary states}

We first want to focus on the stationary states of the system and show that it is characterized by one stable and several unstable stationary states. Even though the unstable states cannot be observed experimentally, they give a clear picture of the structure of the external potential $V$ in Eq. \eqref{eq-Vext}, which also determines the dynamics of the system, discussed in Sec. \ref{sec-Dynamics}.

For a given set of physical parameters $N^2 {\gamma}_{\rho,z}$, $a/a_d$, $\Delta$ and $\Ns$ the stationary states of the system of interacting dipolar BECs are calculated solving the Hamiltonian equations of motion for vanishing time derivatives in Eq. \eqref{eq-Ham-eq-motion}, i.e. $\dot{q}_{\rho,zj} = 0$,  $\dot{p}_{\rho,zj} = 0$. This results in $p_\rho=p_z\equiv 0$ and a $2\Ns$--dimensional, highly nonlinear system of equations for the generalized coordinates $\qrj, \qzj$ which is solved numerically after providing initial values for $\qrj$ and $\qzj$ ($j=1,\ldots,\Ns$). 

\begin{figure}[t]
 	\includegraphics[angle=-90, width=0.88\columnwidth]{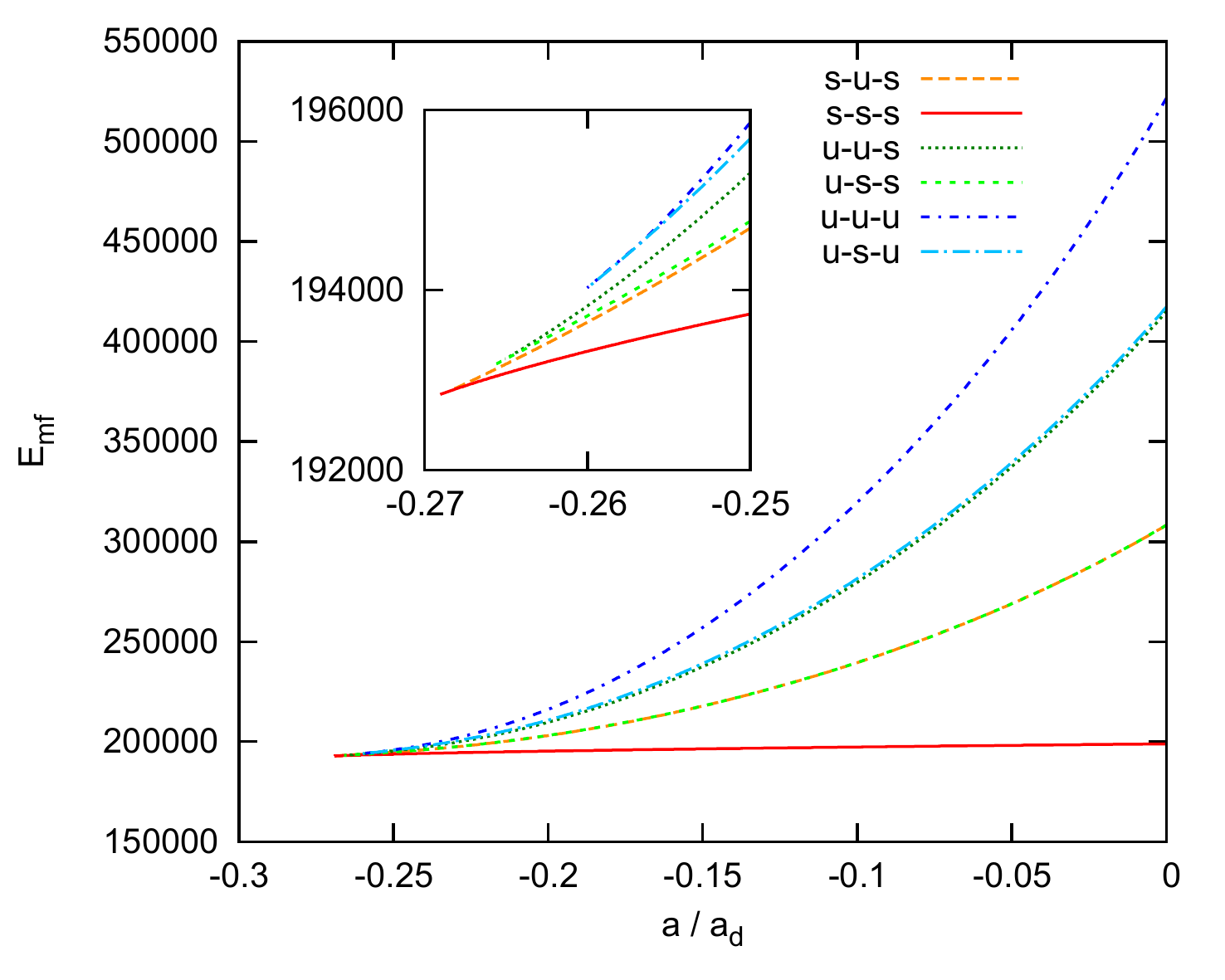}
	\caption{Mean-field energy of the stationary states for a stack of $\Ns=3$ condensates for a distance $\Delta=0.07$, scaled trap frequency $N^2 \bar{\gamma}=1300$ and aspect ratio $\lambda=340$ with $N_b=6$ different branches. Inset: The different branches arise in tangent bifurcations at different values of the scattering length. The symbols ``s'' and ``u'' denote weather of not the individual BECs are stable or unstable.}
\label{Fig-meanfieldenergy}
\end{figure}

For a single BEC there exist two different stationary states for a scattering length above the critical value. One of them is stable (``s'') and one is unstable (``u'') \cite{koeberle2009NJP}. By analogy with that we label the states of the multilayer stack of BECs by a combination of ``s'' and ``u'' meaning that for $\Delta \to \infty$ (i.e. vanishing interaction between the single BECs) the stack would be divided into single BECs, each in a stable or an unstable stationary state.

\begin{figure}[t]
\includegraphics[angle=-90, width=0.88\columnwidth]{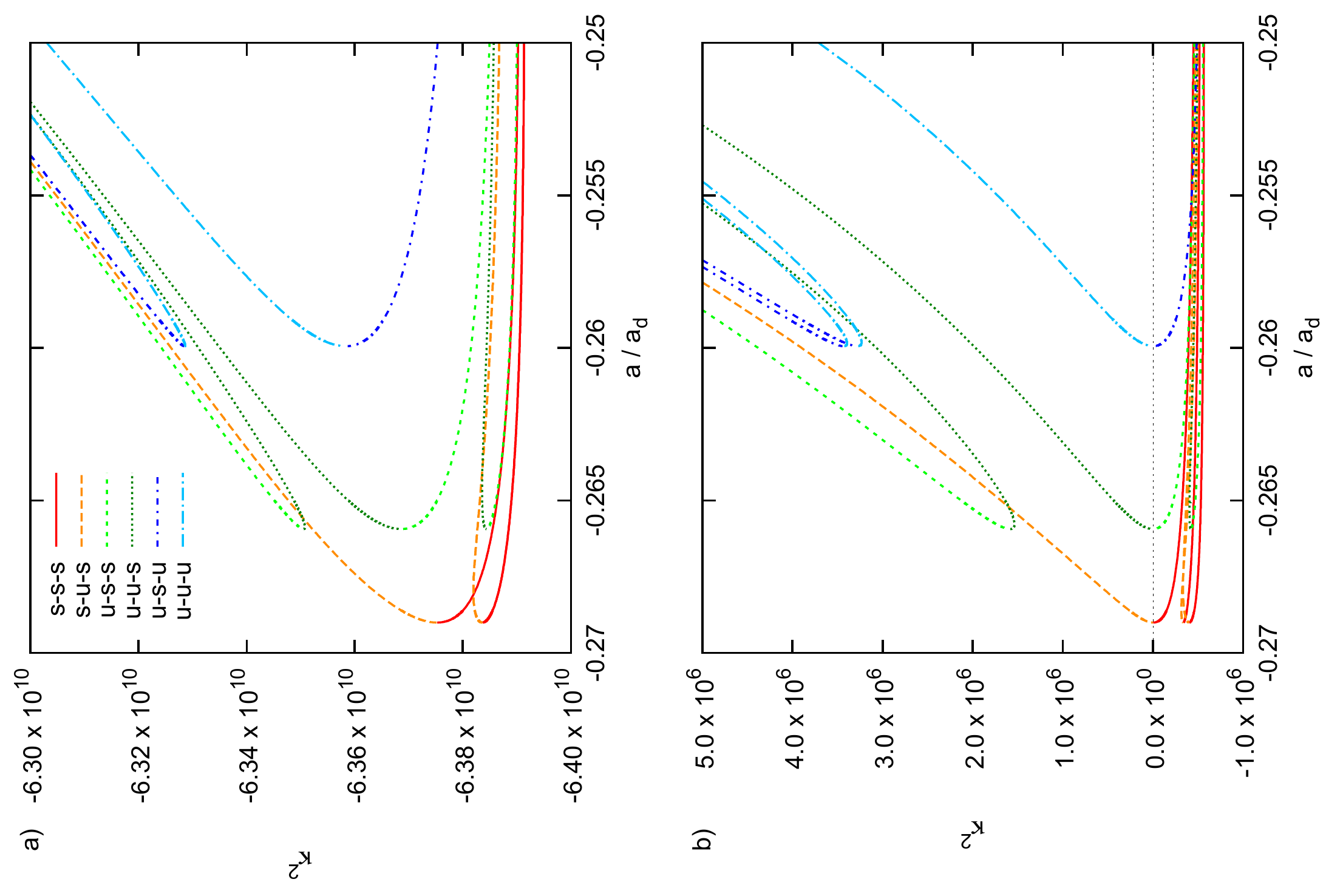}
\caption{Eigenvalues of the Jacobian matrix $J$ for a stack of $\Ns=3$ BECs and different arrangements in dependence on the scattering length. The physical parameters are the same used for the calculation of the mean-field energy in Fig. \ref{Fig-meanfieldenergy}. The state ``s-s-s'' (solid red lines) is the only one with exclusively negative eigenvalues, and in general the eigenvalues corresponding to a motion in $z$--direction (a) differ by several orders of magnitude from those corresponding to the motion in radial direction (b).}
\label{Fig-EigenwerteNs3}
\end{figure}

Fig. \ref{Fig-meanfieldenergy} shows the mean-field energy for stationary states of the stacked BECs for $\Ns=3$ condensates, a distance of $\Delta=0.07$ and a trap geometry defined by $N^2\bar{\gamma} = 1300$ and an aspect ratio of $\lambda=340$ (this trap geometry has been used for calculations in Ref. \cite{koeberle2009} and will also be used for all calculations in this paper). As can be seen there are several branches of the mean-field energy. The number of the stationary states can be explained assuming two different stationary states of each condensate (``s'' and ``u'') which would result in $2^{\Ns}$ possible arrangements for the stack. Since some of the arrangements are physically equivalent, because they only differ by an inversion with respect to the $z=0$--plane, the number of independent different arrangements, and hence the number of branches, is
\begin{equation*}
 	N_b = \begin{cases}
        2^{\Ns/2-1} + 2^{\Ns-1} \qquad 	& \text{for }\Ns \text{ even} \\
	2^{(\Ns-1)/2} + 2^{\Ns-1} 	& \text{for }\Ns \text{ odd.}
       \end{cases}
\end{equation*} The different states arise in tangent bifurcations at different values of the scattering length. We note that two states arising together always belong to the same type of symmetry with respect to the $z=0$--plane: Either both states are symmetric or both are antisymmetric.

The eigenvalues $\kappa^2$ of the Jacobian matrix for the same set of physical parameters used to calculate the mean-field energy are shown in Fig. \ref{Fig-EigenwerteNs3}, and the range of the scattering length corresponds to that of the inset in Fig. \ref{Fig-meanfieldenergy}.  Fig. \ref{Fig-EigenwerteNs3}a and Fig. \ref{Fig-EigenwerteNs3}b show the eigenvalues corresponding to the motion in $\rho$-- and $z$--direction, respectively, which differ from each other by several orders of magnitude because of the large trap aspect ratio of $\lambda=340$. The state labeled ``s-s-s'' is the only one with exclusively negative eigenvalues and consequently is the only stable state. All the other states have at least one positive eigenvalue and are unstable. Note that, for this reason, numerically exact calculations \cite{koeberle2009} can only access the stable ground state but not the excited states. By contrast all stationary states are independently of their stability accessible by the variational approach, which is one of the big advantages of that method.

It is important to note that the stability properties of the multilayer stacks
do not depend on symmetry assumptions for the wave function.
Since the ansatz with a cylindrically symmetrical trial wave functions implies
restrictions, we additionally investigated the stability of the stationary 
states using an extended three-dimensional trial wave function 
\begin{align*}
 \psi_j (&x,y,z,t) = \left( \frac{2^3 \Aixj \Aiyj \Aizj }{ {\pi^3}} \right)^{1/4} \times   \\ 
	\exp& \left( \mathrm{i} A_{x\! j} x^2  + \mathrm{i} A_{y\! j} y^2  +\mathrm{i} A_{z\! j} \! \left(z+ \frac{(\Ns+1-2j)\Delta}{2}  \right)^2 \right), 
\end{align*} which in principle also allows to study anisotropic trap geometries.
It turns out that the stability of the stationary states is not affected by this generalization, in particular the mode that becomes unstable when one goes below the critical scattering length remains cylindrically symmetric. Thus the general features found in the two-dimensional approach remain valid.

We also note that the simple variational ansatz confirms the dependence of an increasing critical scattering length when one adds more condensates to the stack presented in Ref. \cite{koeberle2009}. Moreover it shows that the wave functions of the central condensates are accumulated near the symmetry axes while the outer BECs are more extended in the radial direction.
Of course, as mentioned above, non-Gaussian structures in the wave function
cannot be revealed with this approach.

\subsection{Dynamics}
\label{sec-Dynamics}

\begin{figure}[t]
\includegraphics[angle=-90, width=0.9\columnwidth]{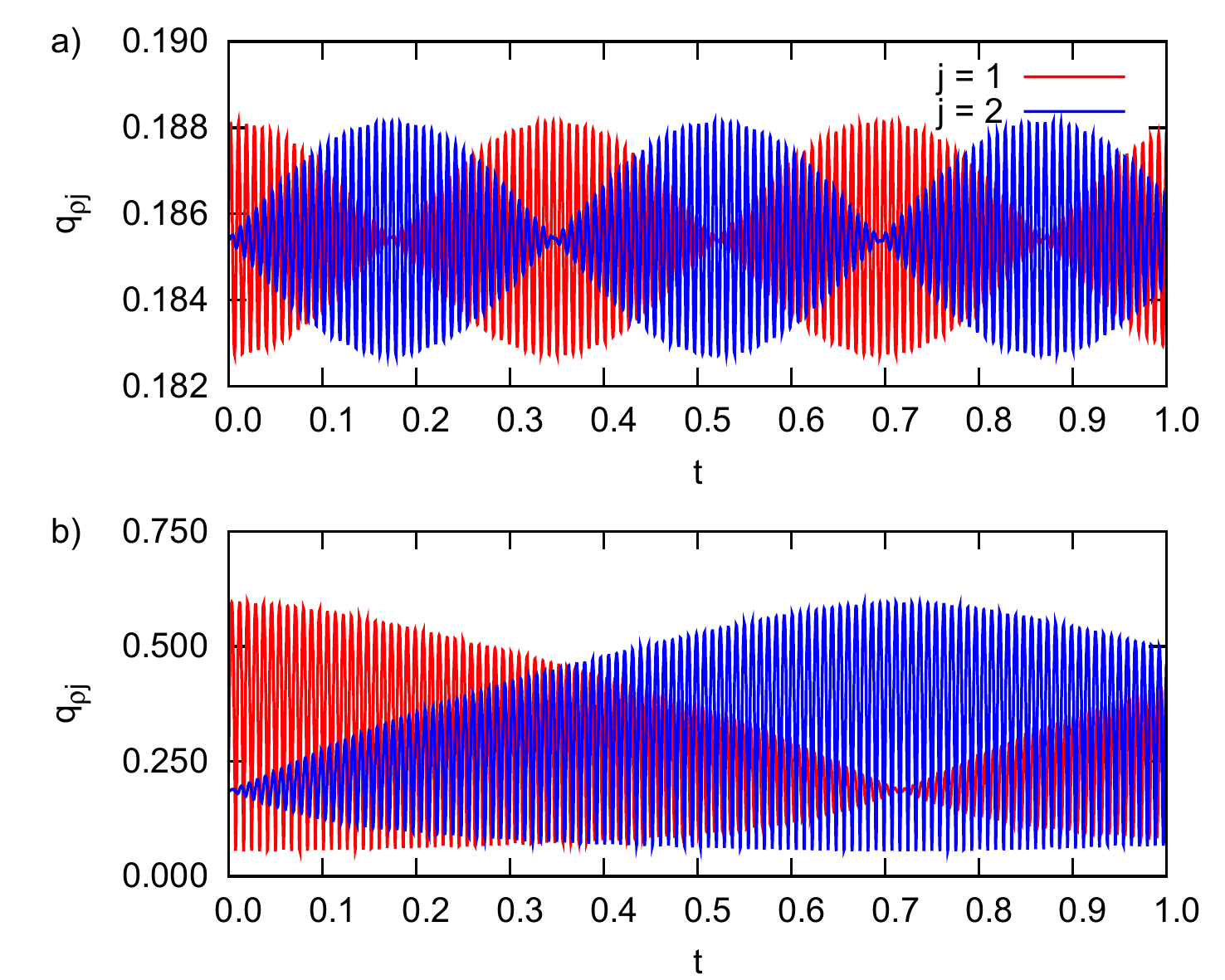}
\caption{Time evolution of a $2$-layer stack of dipolar BECs  described by a particle moving in the external potential \eqref{eq-Vext} for $\Delta=0.035$ and a scattering length of $a/a_d=-0.1$. At $t=0$, the particle is placed in the minimum of the potential $V$ with nonzero initial momentum $p_{\!\rho 1} \neq 0$. The two condensates represented by the particle exchange their excitation energy periodically. For a small initial momentum $p_{\!\rho 1}$ (a) the energy exchange happens in a shorter period of time than it does for large values of the initial momentum (b).}
\label{Fig-Startimpulse-Ns2}
\end{figure}

\begin{figure}[b]
	\includegraphics[width=0.9\columnwidth]{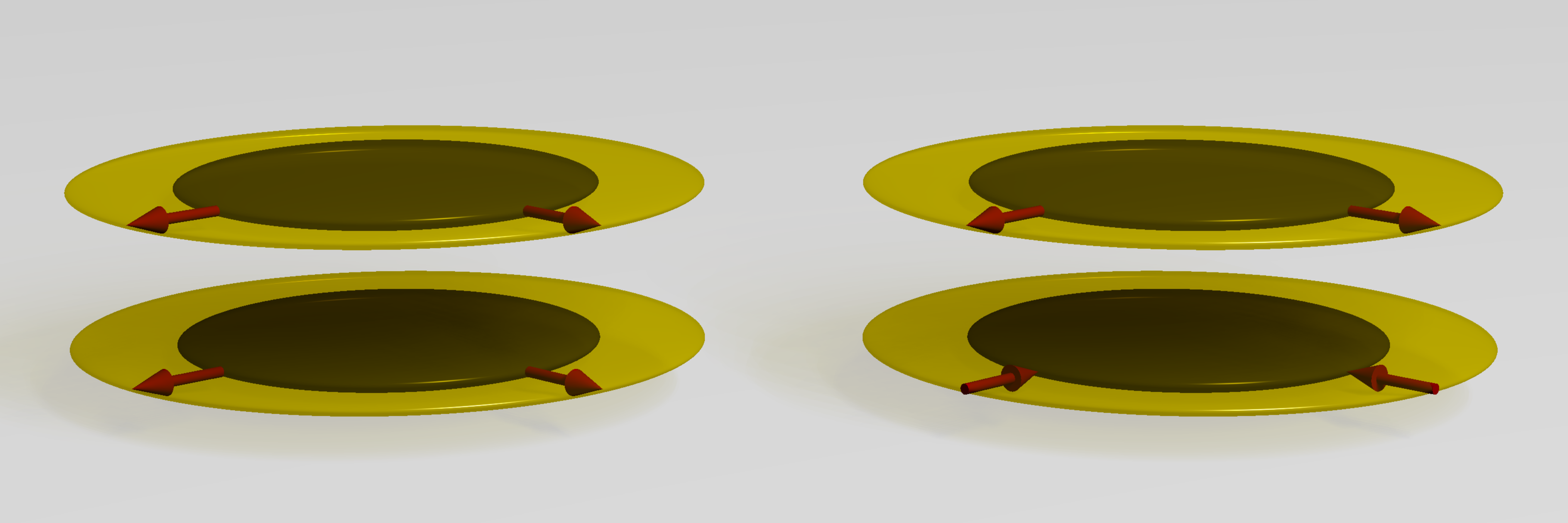}
	\caption{Normal modes of the coupled radial motion of two interacting BECs. The two BECs either oscillate in phase (on the left) or with a phase shift of $\pi$ (on the right) and different oscillation frequencies corresponding to the different normal modes.}
\label{Fig-Eigenmoden}
\end{figure}

We demonstrate the coupled motion of the BECs by placing the particle representing the condensate wave functions at the stable fixed point of the Hamiltonian equations of motion and adding some kinetic energy to the condensate $j=1$ (achieved by the initial condition $p_{\rho 1} \neq 0$, which means the excitation of BEC $j=1$). Fig. \ref{Fig-Startimpulse-Ns2}a shows the time evolution of the radial extension of the wave functions in a stack of two BECs (represented by the $q_\rho$--coordinates of the particle) for this situation with a small value of the initial momentum  $p_{\! \rho 1}$. It is calculated solving the Hamiltonian equations of motion \eqref{eq-Ham-eq-motion} using a Runge-Kutta algorithm. As can be seen, the extension of the excited BEC ($j=1$) begins to oscillate around its stationary value quickly. Due to the interaction, its excitation energy is transferred to the condensate $j=2$ on a larger time scale, and the two BECs continue exchanging energy periodically. The energy exchange can also be observed for more than two BECs, the difference being that the whole energy is not exchanged between two single BECs but transferred to all the others in the stack.

\begin{figure}[t]
\includegraphics[width=0.9\columnwidth]{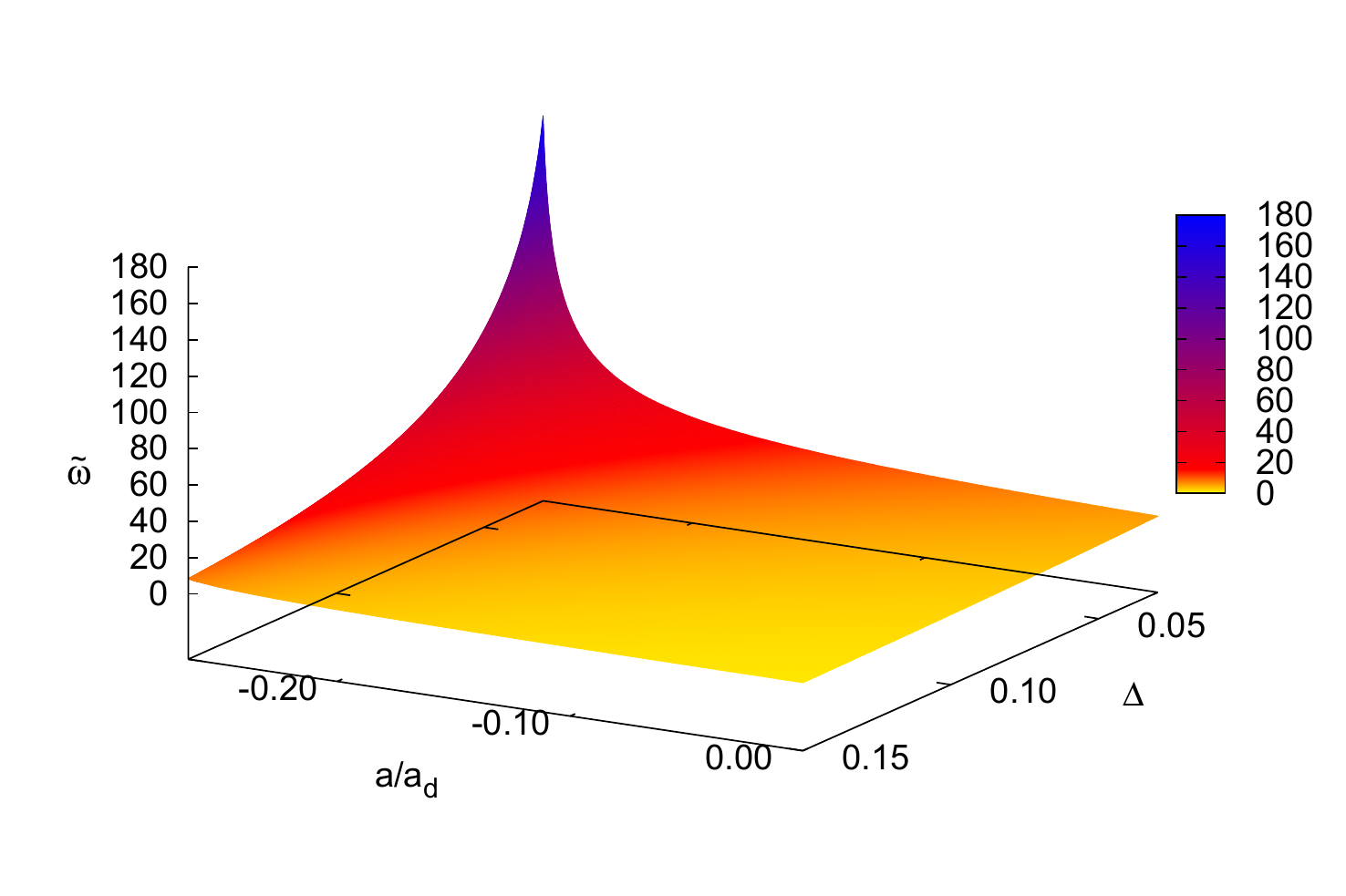}
\caption{The frequency  $\tilde{\omega} = \abs{\omega_1 - \omega_2}/2$ that can be interpreted as a characteristic frequency for the energy exchange between two condensates of Fig. \ref{Fig-Startimpulse-Ns2}. The frequency depends crucially on the distance $\Delta$ between the condensates and the scattering length $a/a_d$ and reaches its highest value for small distances $\Delta$ and $a \to a_\text{cr}$. }
\label{Fig-Schwebungsfrequenz}
\end{figure}

For small excitations the coupled motion of the BECs can be described by linearized equations of motion, which reveal normal modes that show a coupling between the radial motion of all BECs and the motion in $z$--direction, respectively,  caused by the long-range DDI. An investigation of these normal modes and the corresponding eigenfrequencies shows that the energy exchange significantly depends on the scattering length and the distance between the condensates. We demonstrate this for a 2-layer stack of BECs: In this case the linearized motion can be described by one normal mode where the BECs oscillate in phase with a frequency $\omega_1$ and another with a phase shift of $\pi$ and a frequency $\omega_2$ (see Fig. \ref{Fig-Eigenmoden}). Thus, for initial conditions $u_{\rho 1}=0$, ${u}_{\rho 2}\neq 0$ and $\dot{u}_{\rho 1,2}= 0$ the deviation from the fixed point is described by 
\begin{align*}
  	\begin{pmatrix}
	u_{\rho 1} \\ u_{\rho 2}
	\end{pmatrix} &\sim \begin{pmatrix} \sin(\omega_1t)+\sin(\omega_2t) \\%[1ex]
 \sin(\omega_1t)-\sin(\omega_2t) \end{pmatrix}  \\
	&= 2 \cdot  \begin{pmatrix} \sin \left( \frac{\omega_1 + \omega_2}{2}  t \right) \cos\left( \frac{\omega_1 - \omega_2}{2}  t \right) \\[2pt]
 \cos \left(  \frac{\omega_1 + \omega_2}{2}  t \right) \sin \left( \frac{\omega_1 - \omega_2}{2}  t \right) \end{pmatrix},
\end{align*} 
where $\tilde{\omega} = \abs{\omega_1 - \omega_2}/2$ is the frequency of the envelope and can be interpreted as the characteristic frequency of the energy exchange. Fig. \ref{Fig-Schwebungsfrequenz} shows the frequency $\tilde{\omega}$ in dependence on the scattering length $a/a_d$ and the distance $\Delta$ between the condensates. It reaches its highest value for small distances and a scattering length near the critical value. Increasing the distance as well as increasing the scattering length, $\tilde{\omega}$ becomes smaller and vanishes for $\Delta \to \infty$ and  $a/a_d \to \infty$, respectively. The range of frequency shown in Fig. \ref{Fig-Schwebungsfrequenz} reaches from about $\tilde{\omega}=1.4$ to $\tilde{\omega}=180$ which, considering the particle number scaling, corresponds to about $0.5\,$Hz to $70\,$Hz so that this effect of energy exchange should be observable in actual experiments.

\begin{figure}[t]
\includegraphics[width=0.9\columnwidth]{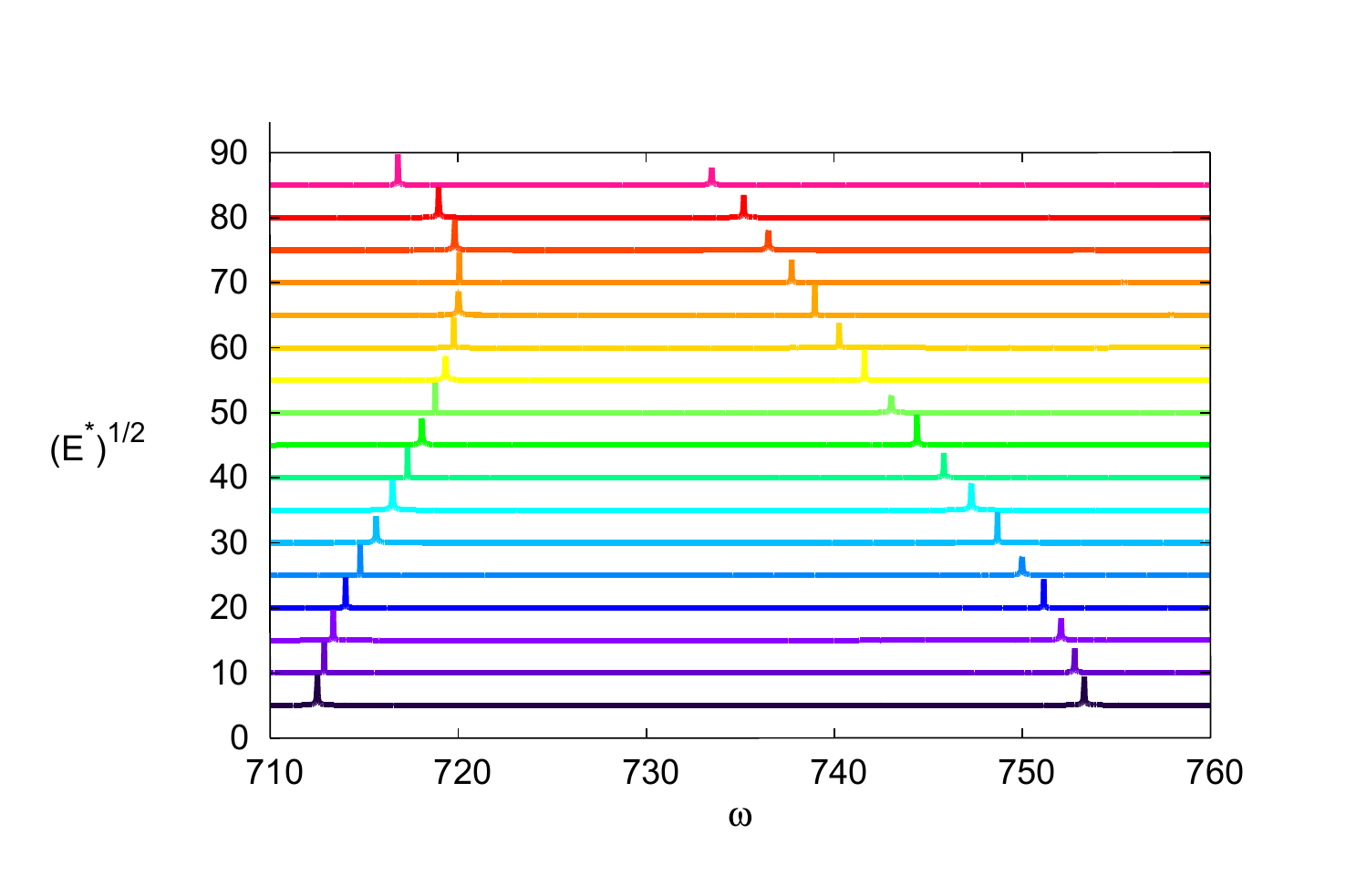}
\caption{Frequency spectra of the radial oscillation of the single condensates in the stack of 2 BECs of Fig. \ref{Fig-Startimpulse-Ns2} for different excitation energies $E^*\!$. The single peaks indicate the fundamental frequencies, and the peak heights are plotted in arbitrary units.}
\label{Fig-Freq-nichtlinear}
\end{figure}

Our calculations with the exact Hamiltonian equations of motion confirm this behavior also for high excitations and additionally reveal a dependence of the frequency of energy exchange on the excitation energy. In case of high excitations the oscillations are anharmonic (see Fig. \ref{Fig-Startimpulse-Ns2}b) and we determine the oscillation frequencies by Fourier transforming the time-dependent extension of the BECs. Fig. \ref{Fig-Freq-nichtlinear} shows the fundamental frequencies of the oscillations for different excitation energies $E^*\!$. Additionally there appear higher harmonics (not shown) whose amplitudes grow with increasing excitation. The frequency of the envelope, however, remains determined solely by the difference of the two fundamental frequencies, which becomes smaller with increasing excitation energy of the stack, i.e. in a highly excited stack the exchange of energy between the single BECs takes longer than it does in slightly excited stacks.

We note that, because of the strong confinement in the $z$--direction, the extension of the BECs in this direction remains small compared to the distance $\Delta$ also for high excitations of the stack so that tunnelling between the single condensates can still be neglected.

\begin{figure}[t]
\includegraphics[angle=-90, width=0.9\columnwidth]{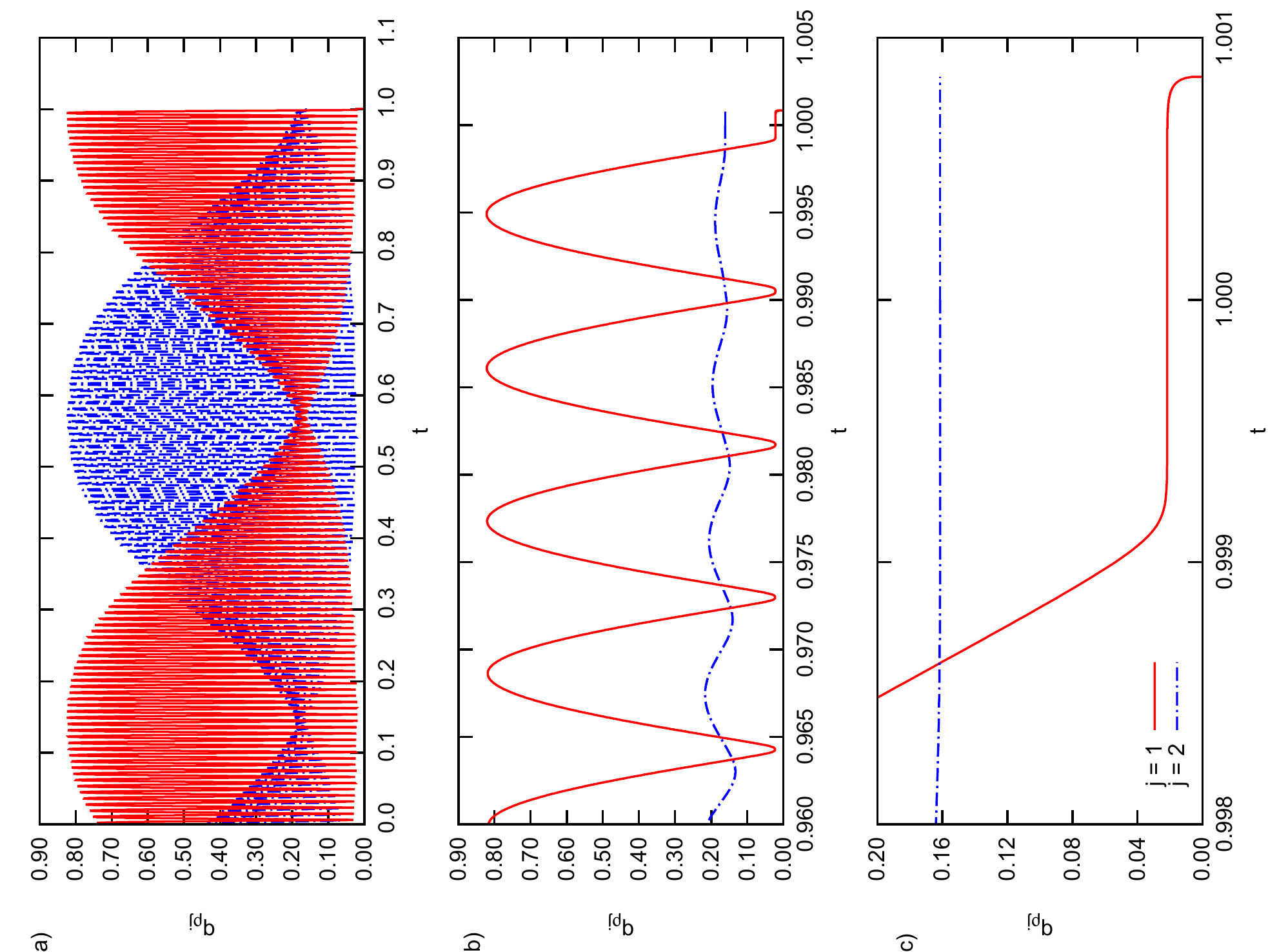}
\caption{Time evolution of the radial extension of the two BECs of Fig. \ref{Fig-Startimpulse-Ns2} described by a particle moving in the external potential $V$ with initial momenta $p_{\!\rho\! j}, p_{\!z\! j} \neq 0$ and initial values for the generalized coordinates that are not the fixed points of the Hamiltonian equations of motion. The mean-field energy of the system lies slightly above the saddle point of the external potential. a) The extension of the BECs oscillates and they exchange their excitation energy. b) The last few oscillations. c) The second BEC ($j=2$) has transferred its whole kinetic energy to the first one which is now highly excited and whose extension finally reaches $q_{\rho 1}=0$, meaning the collapse of the condensate.}
\label{Fig-Uebertragskollaps}
\end{figure}

This exchange of energy can have drastic consequences on the stack of dipolar BECs. We demonstrate this by considering two excited BECs. If the condensates are separated and their energy is below the saddle point energy of the corresponding external potential $V$ the extension of the two BECs will oscillate around its stationary value for all times. The situation is different if these two condensates are placed in a stack where they interact with each other. We show this by exciting two BECs in such a way that the energy of each individual BEC is below the saddle point energy of $V$ but the excitation energy of the whole stack lies slightly above one of the saddle points of $V\!$. Again, the extensions of both condensates oscillate around their stationary values quickly and the condensates exchange energy (see Fig. \ref{Fig-Uebertragskollaps}a). The last few oscillations (see Fig. \ref{Fig-Uebertragskollaps}b) show a highly excited BEC ($j=1$) while the other one ($j=2$) loses its energy and settles down in its stationary state. Fig. \ref{Fig-Uebertragskollaps}c shows the BEC with $j=2$ that has transferred its whole excitation energy onto the first one and resides in its stationary state. The whole excitation energy is now located in the first condensate whose extension oscillates strongly, until at a certain time ($t\approx 1.0$) it becomes so small that the attractive contact interaction ($a/a_d<0$) which is proportional to the density $\abs{\psi(\vec{r})}^2$ becomes dominant and the radial extension of the BEC reaches $q_{\rho 1}\to 0$, meaning its collapse. In the Hamiltonian picture, the particle representing the stack of BECs crosses a saddle point of the external potential $V$ and falls down at the other side. Consequently, excited BECs are able to interact with each other in a way that induces the collapse of one of the BECs in the stack. Of course this behavior can also be observed for more BECs in a stack.

\begin{figure}[t]
\includegraphics[width=0.9\columnwidth]{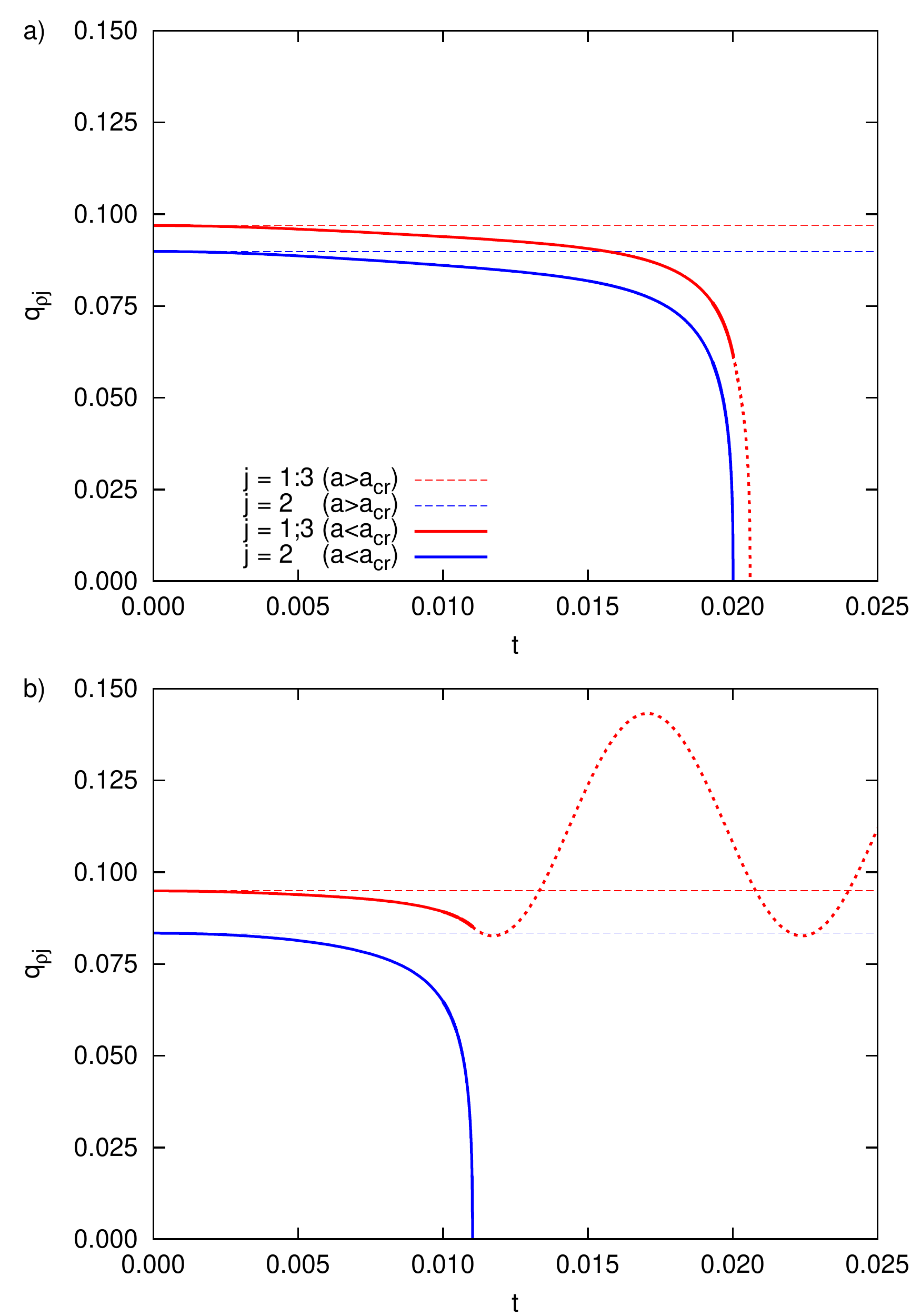}
\caption{Time evolution of a stack of $\Ns=3$ BECs. The multilayer stack of dipolar BECs is assumed to be in the ground state at a scattering length that lies slightly above the critical value and does therefore not change in time (dashed lines). Reducing the scattering length below the critical value, all coordinates $\qrj$ begin to shrink, representing the contraction of the BECs. a) The central BEC (blue line) collapses first but at a distance of $\Delta=0.035$, this also causes the collapse of the other BECs (dotted line). b) The same situation with a distance of $\Delta=0.07$. The central condensate again collapses, but the coupling is not strong enough to determine the other BECs to collapse. }
\label{Fig-MittlererKollaps}
\end{figure}

Finally, we wish to investigate the dynamics of the multilayer stack of BECs when the scattering length is reduced below the critical value. Above that value, there exist several stationary states of the coupled BECs, and we assume the stack to be in the stable ground state at a scattering length that lies slightly above the critical value. If we now decrease the scattering length below $a_\text{cr}$ there no longer exists a stationary state and the resulting dynamics of the stack is shown in Fig. \ref{Fig-MittlererKollaps}a for $\Delta=0.035$ for a stack with 3 BECs. The radial ``extensions'' $\qrj$ of all BECs begin to shrink until the central BEC ($j=2$) collapses ($q_{\rho 2} =0$). The outer two BECs still have a finite extension, but are strongly affected by the collapse of the central condensate. The equations of motion are solved for the two remaining BECs for the time after the collapse of the central BEC (dotted lines) and one can see that the other two BECs will also reach $q_\rho =0$, meaning its collapse. Doubling the distance $\Delta$ between the BECs, the interaction becomes weaker. It is again the central BEC that collapses first, but the effect on its neighbors is not strong enough to force them to collapse, too (see Fig. \ref{Fig-MittlererKollaps}b).

In the framework of this variational approach, three-particle collisions, causing particle losses, have been neglected, but will inevitably happen during the collapse as the density becomes higher. However, we do not expect qualitative changes in the dynamic since the process of considering a constant particle number in the BEC during the collapse and afterwards neglecting its influence completely (as done here) will then only be changed by a continuous decline of the particle number in the central BEC. Nevertheless, we will investigate this by both, variational and numerically exact grid calculations, taking particle losses into account.

\section{Conclusion and outlook}

Describing the multilayer stacks of dipolar BECs variationally we were able to show that such a stack is characterized by several stationary states that arise in tangent bifurcations at different values of the scattering length. Moreover, we could demonstrate coupled dynamics of the stack which reaches from normal modes for small excitations of the stack and an energy exchange between single BECs on experimentally accessible time scales to the induced collapse of a BEC for high excitations. Generally in an excited stack of interacting BECs the individual condensates always exchange energy and this exchange of energy is significantly affected by the scattering length and the distance between the single BECs.

Of course the ansatz of a single Gaussian trial wave function implies restrictions concerning the structure of the wave functions of the individual BECs, and even though it is not capable of reproducing symmetry breaking angular collapse mechanisms, its advantage has to be seen in the description of the dynamics of the stack which is easily accessible. To additionally investigate effects beyond the Gaussian form of the wave function the ansatz can be extended to a Gaussian wave packet for each BEC in the stack. In the case of a single BEC, this extended ansatz is able to reproduce the numerical results \cite{StefanPatrick2010} what can also be expected for the multilayer stack of BECs. Moreover the results found in this paper can serve as a useful guide for investigations of the dynamical effects in the stack of interacting BECs by exact numerical calculations.

%\bibliography{paper}
%Merlin.mbs v4.21 2009-07-09.
%

\end{document}